\documentclass[aps,showpacs,twocolumn,pra]{revtex4-1}
\usepackage{graphicx,lipsum}
\usepackage{subfigure}
\usepackage{amsmath,color}
\def\gappeq{\mathrel{ \rlap{\raise.5ex\hbox{$>$}}
                      {\lower.5ex\hbox{$\sim$}} } }
\def\lappeq{\mathrel{ \rlap{\raise.5ex\hbox{$<$}}
                      {\lower.5ex\hbox{$\sim$}} } }
\usepackage{bm}
\usepackage{color}
\usepackage{epstopdf}

\newcommand{\del}[1]{\textcolor{red}{}}

\graphicspath{{figs/}}

\begin{document}

\title{Engineering Bright Matter-Wave Solitons of Dipolar Condensates}

\author{M. J. Edmonds}
\author{T. Bland}
\author{R. Doran}
\author{N. G. Parker}
\affiliation{Joint Quantum Centre (JQC) Durham--Newcastle, School of Mathematics and Statistics, Newcastle University, Newcastle upon Tyne, NE1 7RU, United Kingdom}

\date{\today{}}

\begin{abstract}\noindent
We present a comprehensive analysis of the form and interaction of dipolar bright solitons across the full parameter space afforded by dipolar Bose-Einstein condensates, revealing the rich behaviour introduced by the non-local nonlinearity. Working within an effective one-dimensional description, we map out the existence of the soliton solutions and show three collisional regimes: free collisions, bound state formation and soliton fusion. Finally, we examine the solitons in their full three-dimensional form through a variational approach; along with regimes of instability to collapse and runaway expansion, we identify regimes of stability which are accessible to current experiments.
\end{abstract}
\maketitle

\section{Introduction}
Solitons have been observed in a profusion of physical systems, and fundamentally exhibit a dual nature possessing both wave and particle like qualities. All solitons owe their existence to the balance between their kinetic and interaction energies, which gives them their defining characteristic of maintaining their form over great distances. Soliton's dual wave-particle nature has led to them being extolled as information carriers in optical systems \cite{mollenauer_2006}, as well as being of fundamental interest as solutions to nonlinear systems \cite{dauxois_2006}.

The realization of Bose-Einstein condensates (BECs) formed from ultracold atomic gases has, over the last two decades, opened a new chapter in nonlinear physics, allowing unprecedented experimental and theoretical insight into quantum mechanical effects of these many-particle systems \cite{bloch_2008}. In the weakly-interacting, low-temperature limit the intrinsic nonlinear dynamics of the condensate are well-described by a mean-field wavefunction that encompasses the behavior of the coherent matter-wave.

At the mean-field level, the van der Waals interactions between particles gives rise to a local nonlinearity that can support soliton solutions in quasi-one-dimensional waveguides. Under repulsive van der Waals interactions, dark solitons (non-dispersive waves of depleted density) are supported in the condensate \cite{frantzeskakis_2010}. Meanwhile, under attractive van der Waals interactions, bright solitons (droplets which are self-bound along the axis of the waveguide) have been observed \cite{khaykovich_2002,strecker_2002,cornish_2006,medley_2014,mcdonald_2014,lepoutre_2016}. Subsequent experiments have explored the effect of collisions with barriers \cite{marchant_2013,marchant_2016}, and also the role of the relative phase for two trapped solitons \cite{nguyen_2014}. Matter-wave interferometry using bright solitons has also been the focus of recent experimental \cite{mcdonald_2014} and theoretical \cite{helm_2015} interest.   

Powerful analytical tools such as the inverse scattering transform (IST) have allowed the investigation of higher-order solitons \cite{gordon_1983} and the derivation of a particle model for the soliton dynamics and interactions, based on the knowledge of the scattering phase shifts \cite{scharf_1992}. This, in turn, has lead to the identification of regimes of chaotic dynamics for three trapped bright solitons \cite{martin_2007} and the observed frequency shifts of trapped bright solitons in a recent experiment \cite{martin_2016}. Examining collisions between bright solitons introduces an extra parameter, namely the relative phase. While these solitons collide freely in one dimension, numerical simulations \cite{baizakov_2004,parker_2008,parker_2009} have demonstrated that in-phase collisions promote the collapse of the condensate in three dimensions, while a relative phase of $\pi$ suppresses the collapse. Indeed, such $\pi$ phase differences are believed to have existed between experimental bright solitons~\cite{strecker_2002,khawaja_2002,cornish_2006}, critical to the observed stability of these states. A strategy to control the relative phase of the solitons has been proposed \cite{billam_2011}. Attractive nonlinearities can also facilitate molecule-like bound states of two bright solitons \cite{gordon_1983,khawaja_2011,martin_2016}, with these states being sensitive to both the relative phase and velocity of the solitons \cite{khawaja_2011}.        

The creation of condensates with atoms possessing significant magnetic dipole moments - $^{52}$Cr \cite{griesmaier_2005,beaufils_2008}, $^{164}$Dy \cite{lu_2011} as well as $^{160}$Dy, $^{162}$ \cite{tang_2015} and $^{168}$Er \cite{aikawa_2012} - afford a new opportunity to explore the interplay of magnetic effects with the coherent nature of the condensate. The dipole-dipole (DD) interaction is anisotropic and long-ranged, contributing a nonlocal nonlinearity to the mean-field equation of motion for the condensate \cite{lahaye_2009}. Importantly, the relative strength of the local to nonlocal interactions can be precisely tuned using Feshbach resonances, allowing the creation of condensates possessing a dominantly dipolar character \cite{koch_2008}. The recent observation of the Rosensweig instability in a condensate of $^{164}$Dy atoms \cite{kadau_2016} has focused interest towards the manifestation of self-bound droplet phases \cite{barbut_2016,schmitt_2016,chomaz_2016}, where conventionally one would expect the condensate to undergo collapse. The unexpected stability of these states has been attributed to many-body effects \cite{baillie_2016,bisset_2016,wachtler_2016,wachtler_2016a}. There has also recently been the creation of a spin-orbit coupled dipolar degenerate Fermi gas, opening a new route to the study of complex phases in analogy with condensed matter physics \cite{burdick_2016}.

Solitons whose existence depends on a non-local rather that a purely local nonlinearity represent a burgeoning area of interest in many disciplines of physics \cite{rotschild_2006,piccardi_2011,jang_2013}, and dipolar condensates provide a highly-tunable platform to study such solitons. The effect of varying the relative strength of the local and nonlocal interactions has revealed novel bright \cite{cuevas_2009,baizakov_2015,umarov_2016} and dark dipolar matter wave solitons \cite{pawlowski_2015,bland_2015,edmonds_2016,bland_2016} in one dimension. Two-dimensional bright solitons are also predicted to be supported by the anisotropy of the dipole-dipole interaction \cite{pedri_2005,tikhonenkov_2008,raghunandan_2015,chen_2016}. These works highlight important prevailing physical characteristics, including the existence of bound states of multiple dipolar solitons due to the intrinsic long-ranged nature of the dipolar interaction.

We begin by deriving the one-dimensional mean-field equation of motion for the dipolar condensate confined to a quasi-one-dimensional waveguide (Sec.~\ref{sec:tm}). In Sec.~\ref{sec:dbs}, we procure the dipolar bright soliton solutions as a function of the dipole-dipole strength and polarization angle, for attractive and repulsive van der Waals interactions. Subsequently in Sec.~\ref{sec:coll} we focus on the interplay of the relative phase with the dipole-dipole interaction strength on soliton-soliton collisions, highlighting regimes of free collisions, bound state formation and soliton fusion, as well as the effect of noise on such collisions. In Sec.~\ref{sec:stab} we characterize the stability of the bright solitons in three dimensions, revealing the parameter regimes where the solitons are stable to collapse and expansion. Our findings are then summarized in the Conclusion (Sec.~\ref{sec:con}). 

\section{Theoretical Model\label{sec:tm}}
We consider an ensemble of weakly-interacting atoms forming a BEC at zero temperature. Each atom has a mass $m$ and permanent magnetic dipole moment $d$, polarized in a common direction by an external magnetic field. In the dilute limit described by the Gross-Pitaevskii theory, interactions are described by the two-body pseudo-potential
\begin{equation}\label{eqn:pp}
U({\bf r}-{\bf r}')=g\delta({\bf r}-{\bf r}')+U_{\text{dd}}({\bf r}-{\bf r}').
\end{equation}
The first term in Eq.~\eqref{eqn:pp} describes the conventional short-range van der Waals interactions, with $g=4\pi\hbar^2a_s/m$ where $a_s$ defines the $s$-wave scattering length. The second term describes the long-ranged anisotropic dipole-dipole interaction, given by
\begin{equation}\label{eqn:udd}
U_{\text{dd}}(r)=\frac{C_{\text{dd}}}{4\pi}\hat{e}_{j}\hat{e}_{k}\frac{(\delta_{jk}-3\hat{r}_{j}\hat{r}_{k})}{r^3},
\end{equation}
where $C_{\text{dd}}=4\pi d^2$ characterizes the dipole-dipole interaction strength. Here $\hat{e}_j$ defines the unit vector in the direction of the coordinate $\hat{r}_{j}$.  Equivalently one can also show that Eq.~\eqref{eqn:udd} can be written in the form 
\begin{equation}
U_{\rm dd}({\bf r}-{\bf r}')=\frac{C_{\rm dd}}{4\pi}\frac{1-3\cos^{2}\theta}{|{\bf r}-{\bf r}'|^3},
\end{equation}
where $\theta$ defines the angle between the vector joining two dipoles and the direction of polarization. At the magic angle $\theta_{\rm m}\simeq 54^{\circ}$ the dipole-dipole interaction vanishes.  When $\theta<\theta_{\rm m}$ the dipole-dipole interaction is attractive with dipoles lying head-to-tail. When $\theta>\theta_{\rm m}$ the interaction is instead repulsive and the dipoles are orientated in a side-by-side configuration. Rather than defining individually the dipolar and van der Waals interaction strengths, it is convenient to work in terms the ratio $\varepsilon_{\rm dd}=C_{\rm dd}/3g$. It is further possible to consider the parameter region for $C_{\rm dd}<0$ \cite{giovanazzi_2002}, where under rapid rotation the dipoles can be thought of as anti-dipoles, effectively reversing the attractive and repulsive regimes of interaction. The form of the dipole-dipole interaction given by Eq.~\eqref{eqn:udd} will be used to perform the dimensional reduction in momentum space in what follows. The dynamics of the quantum gas are encompassed by the mean-field wave function $\Psi({\bf r},t)$, which defines the atomic density distribution as $n({\bf r},t)=|\Psi({\bf r},t)|^2$. The equation of motion for $\Psi({\bf r},t)$ is given by the dipolar GPE \cite{santos_2000}
\begin{equation}\label{eqn:gpe3d}
i\hbar\frac{\partial\Psi}{\partial t}=\bigg[-\frac{\hbar^2}{2m}\nabla^{2}+\frac{1}{2}m\omega_{\perp}^{2}r_{\perp}^{2}+g|\Psi|^2+\Phi({\bf r},t)\bigg]\Psi.
\end{equation}
The trapping potential appearing in Eq.~\eqref{eqn:gpe3d} has a transverse trapping frequency $\omega_{\perp}$, where $r_{\perp}^{2}=x^2+y^2$ defines the coordinate in the radial direction. As such the potential effectively forms a waveguide, which is uniform along the axial direction, $z$.  Meanwhile, the nonlocal mean-field potential generated by the dipoles appearing in Eq.~\eqref{eqn:gpe3d} takes the form
\begin{equation}\label{eqn:phi3d}
\Phi({\bf r},t)=\int {\rm d}^3{\bf r}'~U_{\text{dd}}({\bf r}-{\bf r}')|\Psi({\bf r}',t)|^2.
\end{equation}
We consider the zero temperature limit of the quasi-one-dimensional dipolar condensate, under which the three-dimensional GPE can be reduced to an equation of motion for the axial degree of freedom. Rigorous analysis of this dimensional reduction has been previously reported \cite{parker_2008a,cai_2010}. The transverse degrees of freedom are assumed to be in their harmonic ground state, which is encapsulated by the condition $\hbar\omega_{\perp}\gg\mu$, where $\mu$ is the 3D chemical potential (i.e. that associated with Eq.~\eqref{eqn:gpe3d} above). The ansatz for the real-space wave function is assumed to be $\Psi({\bf r},t)=\psi_{\perp}(r_{\perp})\psi(z,t)$, where $\psi_{\perp}(r_{\perp})=(l_{\perp}\sqrt{\pi})^{-1}\exp(-r_{\perp}^{2}/2l_{\perp}^{2})$, and $l_{\perp}=\sqrt{\hbar/m\omega_{\perp}}$ defines the transverse harmonic length scale. The Fourier transform of the real-space density is given by $\tilde{n}(k)=\tilde{n}_{z}(k_{z})\exp(-l_{\perp}^{2}(k_{x}^{2}+k_{y}^{2})/4)$, where $k_{i}$ is the momentum associated with coordinate $i$ and $\tilde{n}_{z}(k_z)$ is the momentum space density in the axial direction. To perform the dimensional reduction, we first make use of the mathematical identity \cite{craig_1999}
\begin{equation}\label{eqn:retomom}
\frac{1}{4\pi r^3}\bigg(\delta_{jk}-3\hat{r}_{j}\hat{r}_{k}\bigg)=\frac{2}{3}\delta_{jk}\delta({\bf r})-\delta_{jk}^{\perp}({\bf r}).
\end{equation}
Equation \eqref{eqn:retomom} introduces the quantity $\delta_{jk}^{\perp}({\bf r})$ which gives the transverse part of the $\delta$-function, defined as
\begin{equation}\label{eqn:tdelta}
\delta_{jk}^{\perp}({\bf r})=\int\frac{{\rm d}^{3}{\bf k}}{(2\pi)^3}(\delta_{jk}-\hat{k}_{j}\hat{k}_{k}),
\end{equation}
for unit vector in momentum space, $\hat{k}_{j}$. Then, along with the Fourier representation of the $\delta$-function, one can write down the Fourier transform of Eq.~\eqref{eqn:udd} as
\begin{equation}\label{eqn:u3dk}
\tilde{U}_{\rm dd}({\bf k})=\frac{C_{\rm dd}}{3}\hat{e}_{j}\hat{e}_{k}(3\hat{k}_{j}\hat{k}_{k}-\delta_{jk}).
\end{equation}
The one-dimensional analogue of Eq.~\eqref{eqn:phi3d} can then be found using the definition given by Eq.~\eqref{eqn:u3dk} along with the momentum space form of the density ${\tilde{n}}(k)={\tilde{n}}_{z}(k_z)\exp(-l_{\perp}^{2}(k_{x}^{2}+k_{y}^{2})/4)$. This gives
\begin{equation}\label{eqn:u1d}
\frac{{\tilde{U}}_{\rm 1D}(k_z)}{l_{\perp}}=4U_{0}\bigg[\frac{k_{z}^{2}l_{\perp}^{2}}{2}e^{k_{z}^{2}l_{\perp}^{2}/2}E_{1}\bigg(\frac{k_{z}^{2}l_{\perp}^{2}}{2}\bigg)-1\bigg]+\frac{8}{3}U_{0},
\end{equation}
\begin{figure}
\includegraphics[scale=0.9]{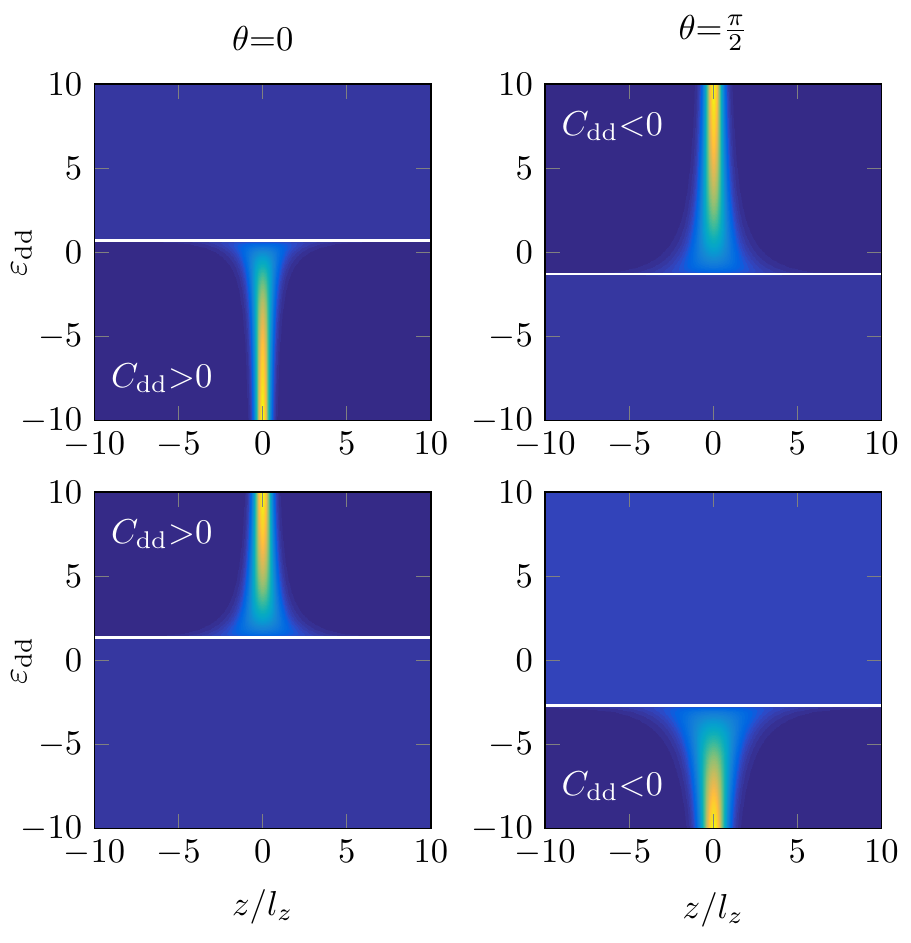}
\caption{\label{fig:carpet} Ground state density profiles obtained numerically from Eq.~\eqref{eqn:gpe1d} as a function of $\varepsilon_{\rm dd}$. The top (bottom) row corresponds to sgn($a_s$)$<0$ (sgn($a_s$)$>0$).  The left (right) column corresponds to $\theta=0$ ($\theta=\pi/2$).  The solid white line in each figure indicates the borderline between homogeneous and soliton solutions. Solutions are for $\sigma=0.2$.}
\end{figure}

\noindent where the parameter $U_{0}=C_{\rm{dd}}[1+3\cos(2\theta)]/(32\pi l_{\perp}^{3})$ encapsulates the strength of the dimensionally-reduced dipole-dipole interaction \cite{giovanazzi_2004}. The momentum associated with the $z$-axis is defined as $\hbar k_z$, while $E_{1}(x)=\int_{x}^{\infty}dt\ t^{-1}e^{-t}$ defines the exponential integral. Then, the one-dimensional mean-field model is encompassed by the pseudo-potential ${\tilde{U}}_{\rm tot}(k_z)=g/(2\pi l_{\perp}^{2})+{\tilde{U}}_{\rm 1D}(k_z)$, which includes the van der Waals as well as dipolar contribution. Equation \eqref{eqn:u1d} will be utilized to find the family of bright soliton solutions to the quasi-one-dimensional dipolar Gross-Pitaevskii equation. The effective 1D dipolar GPE is then written as
\begin{equation}\label{eqn:gpe1d}
i\hbar\frac{\partial\psi}{\partial t}=\bigg[-\frac{\hbar^2}{2m}\partial_{z}^{2}+\frac{g}{2\pi l_{\perp}^{2}}|\psi|^2+\Phi_{\rm{1D}}(z,t)\bigg]\psi,
\end{equation}
where the one-dimensional dipolar potential is obtained via the convolution theorem as $\Phi_{\rm 1D}(z,t)=\mathcal{F}^{-1}[{\tilde{U}}_{\rm 1D}(k_z)\tilde{n}_z(k_z,t)]$, where $\mathcal{F}^{-1}[\dots]$ denotes the inverse Fourier transform.
To obtain the solutions to Eq.~\eqref{eqn:gpe1d} we work in momentum space using a split operator method. In what follows we adopt the so-called `soliton' units \cite{martin_2007}, where length, time and energy are measured in terms of $l_{z}=\hbar/mv$, $\tau=l_z/v$, and $E=mv^2$ respectively, where $v=|g_{\rm 1D}|N/\hbar$ and $g_{\rm 1D}=g/(2\pi l_{\perp}^{2})$ defines the units of velocity and one-dimensional van der Waals interaction strength respectively. We quantify how one-dimensional the system is through the ratio $\sigma=l_{\perp}/l_z$. To be in the true one-dimensional limit one must have $\sigma\ll 1$.
\begin{figure*}[t]
\includegraphics[scale=0.85]{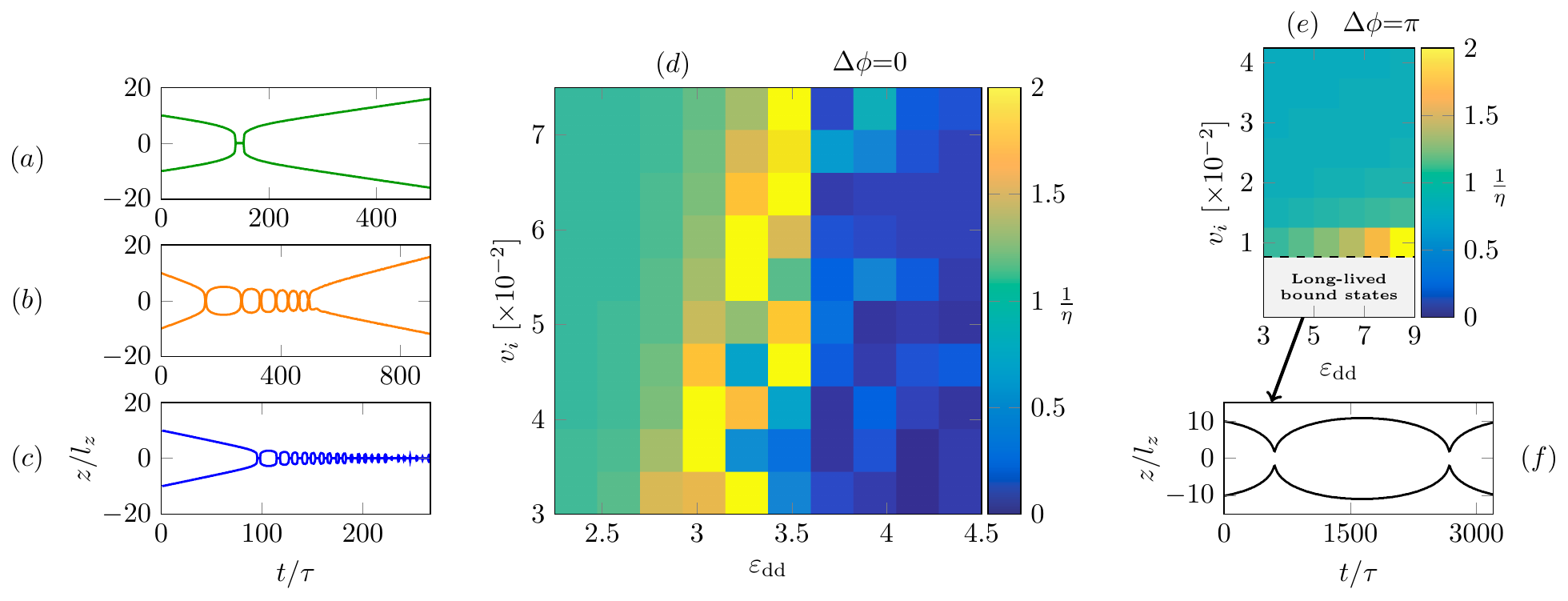}
\caption{\label{fig:swatch}Collision dynamics of in-phase and out-of-phase solitons. For in-phase collisions three regimes of dynamics are shown: (a) elastic, (b) bound-state and (c) inelastic dynamics. The coefficient of restitution (Eq.~\eqref{eqn:cor}) is mapped out as a function of the initial speed $v_i$ and $\varepsilon_{\rm dd}$ in (d). The corresponding restitution data is computed for the out-of-phase collisions in (e).  (f) shows a typical long-lived bound state for out-of-phase dynamics, corresponding to $v_i=5\times 10^{-3}$ and $\varepsilon_{\rm dd}=8$.}
\end{figure*} 
\section{Dipolar Bright Solitons\label{sec:dbs}}

The local cubic nonlinear Schr\"odinger equation (Eq.~\eqref{eqn:gpe1d} in the limit $\varepsilon_{\rm dd}=0$) is known to possess bright soliton solutions for $a_s<0$ (corresponding to when the chemical potential $\mu<0$). The single soliton solution, which for simplicity we take to be initially positioned at the origin is
\begin{align}\nonumber
&\psi(z,t)=\frac{1}{2\sqrt{l_{z}}}\text{sech}\bigg(\frac{z-ut}{2l_{z}}\bigg)\\&\times\exp\bigg[i\bigg\{\frac{m}{\hbar}\bigg(uz{+}\frac{u^2t}{2}{+}\frac{\omega_{\perp}^{2}|a_{s}|^2N^2t}{2}\bigg){+}\phi\bigg\}\bigg]\label{eqn:bsol}
\end{align}  
where $u$ defines the velocity of the bright soliton, while $\phi$ is the phase. This solution describes a sech-shaped profile which propagates at constant velocity. Here, we explore the family of dipolar bright solitons across the full parameter space afforded by the quasi-one-dimensional model. For a single bright soliton, we can independently vary four key parameters: $\varepsilon_{\rm dd}$, $\theta$, sgn($a_s$) and also $\sigma$. Meanwhile the normalization is given by $\int {\rm d}z|\psi(z,t)|^2=N$. Despite the additional dipolar term present in Eq.~\eqref{eqn:gpe1d}, we will see that the allowed dipolar bright soliton solutions are still sech-shaped. Throughout this work we take $\sigma=0.2$.

Figure \ref{fig:carpet} shows the ground state densities obtained by solving Eq.~\eqref{eqn:gpe1d} numerically in imaginary time. Each individual plot is divided into two regions: a flat homogeneous region, corresponding to regimes of net repulsive interactions, and a second region showing the inhomogeneous dipolar bright soliton solutions. The solid white line in each figure shows the crossover between these two parts, which corresponds to when the 1D chemical potential of the ground state solution crosses from positive (homogeneous state) to negative (bright soliton). The top (bottom) rows in Fig.~\ref{fig:carpet} correspond to sgn($a_s$)$<0$ ($>0$). Fixing both the sign and value of the van der Waals interactions reveals that altering the polarization angle of the dipoles between $\theta=0$ and $\theta=\pi/2$ has the effect of shifting the soliton solutions from $C_{\rm dd}>0$ ($\theta=0$) to $C_{\rm dd}<0$ ($\theta=\pi/2$). The parameter regimes where $C_{\rm dd}<0$, corresponding to anti-dipoles are found to support soliton solutions as the net interactions are attractive for dipoles polarized perpendicular to the $z$-axis. We note that in each of the cases presented in Fig.~\ref{fig:carpet} the borderline between the homogeneous and soliton solutions does not coincide with $\varepsilon_{\rm dd}=0$. This can be understood from the form of the dimensionally-reduced pseudo-potential, Eq.~\eqref{eqn:u1d} which comprises both a nonlocal and local contribution, whose net effect is to shift the value of $\varepsilon_{\rm dd}$ at which the chemical potential $\mu$ changes sign.

\section{Collisions\label{sec:coll}}
In the absence of dipolar interactions and within the one-dimensional nonlinear Schr\"odinger equation, bright solitons are known to collide elastically, emerging from the collision with their original speed and form. The net effect of soliton-soliton interaction is a position and phase shift in the outgoing solitons. In this section we study the collisions of two dipolar bright solitons, exploring the role the relative phase plays in collisions as a function of the dipole-dipole interaction and the initial kinetic energy of the solitons. In what follows we simulate two-counter-propagating solitons with equal speed $v_i$, and take sgn$(a_s)=1$ and $C_{\rm dd}>0$, i.e. the soliton solutions shown in the lower left figure of Fig.~\ref{fig:carpet}.
\subsection{In-phase collisions}
We consider the collision dynamics of bright solitons with zero initial phase difference, $\Delta\phi=0$. As we shall see, the collisions of the dipolar solitons can be inelastic. In order to quantify the elasticity of the soliton dynamics, we compute the coefficient of restitution, defined as
\begin{equation}\label{eqn:cor}
\eta=\frac{v_{1}(t_{f})-v_{2}(t_{f})}{v_{1}(t_{i})-v_{2}(t_{i})}
\end{equation}
where $v_{j}$ is the velocity of soliton $j$, and $t_i$ and $t_f$ are the initial and final times, respectively. The coefficient of restitution is a measure of the amount of kinetic energy before and after a collision event. If $\eta=1$ the incoming and outgoing speeds are identical and a collision is perfectly elastic.   For $\eta<1$ the outgoing speeds are less than the incoming speeds, and the collision is deemed inelastic. As we shall see it is also possible to realize $\eta>1$, corresponding to the outgoing speeds exceeding the incoming speeds, and which can occur, for example, when interaction (van der Waals plus dipolar) energy is transformed into kinetic energy during the collision. 

The coefficient of restitution $\eta$ is mapped out in the ($\varepsilon_{\rm dd}$,$v_i$) plane in Fig.~\ref{fig:swatch} (d). Each pixel represents an individual simulated collision between two bright solitons, with the color representing the value of the inverse of the coefficient of restitution (we plot $1/\eta$ as this quantity's scale evolves at a slower rate over the parameter range considered). Three different regimes of dynamics can then be identified. For relatively weak dipole-dipole interactions $\varepsilon_{\rm dd}\leq 2.5$, the collisions are almost perfectly elastic ($\eta\sim1$), independent of the initial velocity. Figure~\ref{fig:swatch} (a) shows a typical set of collision trajectories in this limit. Here the incoming solitons scatter elastically with each other and escape at longer times. In the intermediate regime, short-lived (meta-stable) bound states are found, whose dynamics are inelastic. Here, the balance of the initial kinetic energy of the solitons to the dipole-dipole strength is favorable to the formation of a short-lived bound state; again a simulation typical of this situation is shown in Fig.~\ref{fig:swatch} (b). We note that similar dynamics were reported by Ref.~\cite{campbell_1986} for kink solutions to the sine-Gordon equation, which also exhibited short-lived bound-states. In contrast to this work, we observe a single transition from bound state to free solitons, as the speed of the incoming soliton is increased. In the limit where the dipole-dipole interactions are large, we instead observe soliton fusion with $1/\eta\ll 1$. A trajectory plot indicative of this limit is shown in Fig.~\ref{fig:swatch} (c), showing the collision and eventual merging of the two solitons. Here the individual solitons do not re-emerge at long times.

We can gain an understanding of the short-lived bound state by considering the effect of the in-phase collision on the soliton's kinetic and potential energies. After the bound state has initially formed, each successive collision event causes a redistribution of some interaction (van der Waals and dipolar) energy into kinetic energy, causing the effective oscillation frequency of the soliton molecule to increase. Eventually, this is enough to cause the bind to break, releasing the two solitons.  Note that the total energy remains constant throughout the simulations. 

\begin{figure}[t]
\includegraphics[scale=0.7]{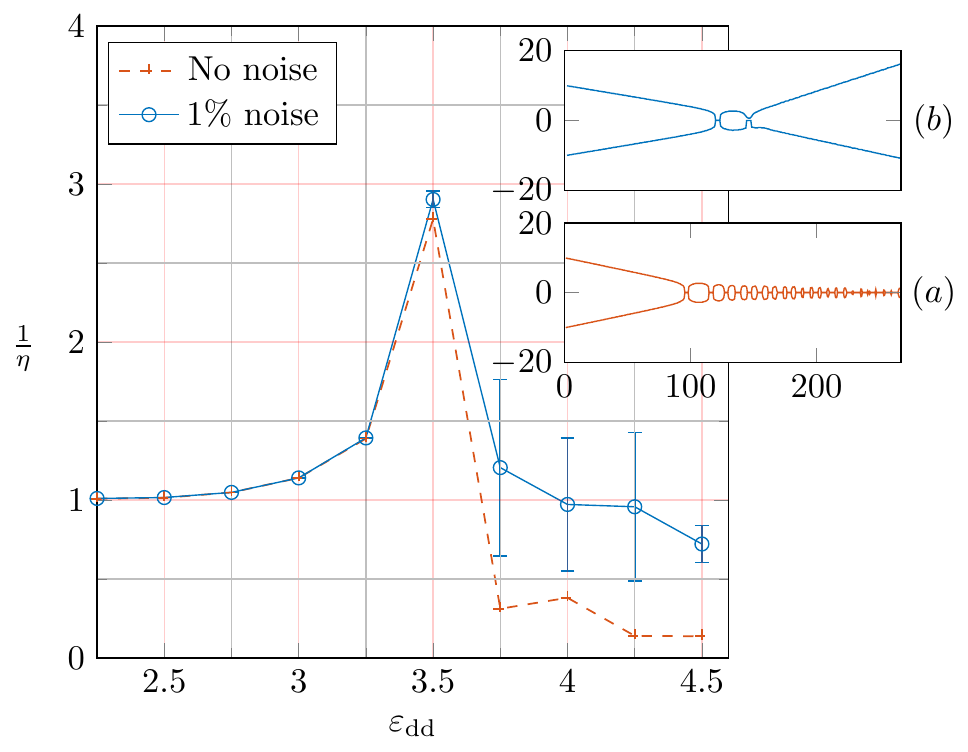}
\caption{\label{fig:noise} Comparison of $1/\eta$ with and without noise for in-phase soliton collisions. The initial velocity in all simulations was $v_i=7.5\times 10^{-2}$, corresponding to a line scan along the top row of Fig.~\ref{fig:swatch} (d). The two insets (a) and (b) show trajectories associated with $\varepsilon_{\rm dd}=4.5$ without and with noise respectively.}
\end{figure}

\subsection{Out-of-phase collisions}
It is also pertinent to consider the equivalent dynamics for out-of-phase dipolar bright solitons, with Fig.~\ref{fig:swatch} (e) plotting $1/\eta$ in the ($\varepsilon_{\rm dd}$,$v_i$) plane.  For relatively large incoming speed, the dynamics are almost elastic $(\eta\sim 1)$, while for increasing dipole strength or decreasing incoming velocity, the dynamics are instead found to be increasingly inelastic. This system has a regime of long-lived bound states occurring for low incoming speeds, with example collision trajectories shown in the panel Fig.~\ref{fig:swatch} (f).  

The binding of two out-of-phase dipolar bright solitons, has been studied previously by Refs.~\cite{umarov_2016,baizakov_2015}. Unlike their in-phase counterparts, the $\pi$ phase difference preserves long-lived bound states.  One cannot assign a value of $\eta$ to the collisions in this regime.

\subsection{Noise}
In order to quantify the collisional sensitivity of the dipolar bright solitons, we calculate the coefficient of restitution in the presence of noise. The noise is implemented by introducing a random term to the phase with mean zero whose amplitude $\mathcal{N}_{\rm noise}$ is a given fraction of the initial peak density of the soliton density such that
\begin{equation}\label{eqn:noise}
\mathcal{N}_{\rm noise}=\mathcal{N}_{0}\text{max}(n(z)),
\end{equation}
where $0<\mathcal{N}_{0}\leq 1$, and max($n(z)$) is the peak soliton density. As an example we consider $1/\eta$ versus $\varepsilon_{\rm dd}$, for a fixed incoming speed and for in-phase collisions. Figure \ref{fig:noise} shows a comparison of $1/\eta$ with $\mathcal{N}_{0}=0$ (no noise) and $\mathcal{N}_{0}=10^{-2}$. In the presence of noise, each data point was obtained by averaging over 10 individual simulations. The error bars represent the standard deviation.  For low values of $\varepsilon_{\rm dd}$, the value of $1/\eta$ follows very closely the value obtained in the absence of noise. This is attributed to the elastic dynamics being insensitive to the phase noise. On the other hand, for stronger dipolar interactions, there is a marked deviation from the no-noise case, resulting in a larger value of $1/\eta$. In this regime, the presence of noise introduces an apparent repulsion between the solitons, which is large enough to make the collisions shown in Fig.~\ref{fig:noise} elastic. Such an effect has also also been noted in non-dipolar bright soliton collisions \cite{wuster_2009}. We see qualitatively the same behaviour for different strengths of noise, with increasing amounts of noise causing the bound-states to break sooner, until no bound-states are formed.

A corollary of the phase noise is that the bound state dynamics in the presence of noise show fewer oscillations before escaping their bind. This effect is illustrated in the soliton trajectories with/without noise in insets Fig.~\ref{fig:noise} (a) and (b). 
The presence of only small amounts of noise demonstrates how sensitive the binding dynamics are: Fig.~\ref{fig:noise} (b) illustrates that only one oscillation can occur in this example before the solitons escape. For larger amounts of noise the bound states are no longer present, even at these larger values of $\varepsilon_{\rm dd}$. On the other hand, the collsions for out-of-phase solitons, in contrast, are insensitve to noise.  The effective repulsion in these collisions serves to stabilize the collisions against the noise.

Although the analysis presented here is rudimentary, it nonetheless allows one to comment on the effect of dissipative processes that are present in a real system, especially those at finite temperature. For example, for small changes in temperature, it is expected that the formation of bound states would be unfavorable. The dipolar interactions would play an increasingly diminished role in the system dynamics, since this term fundamentally depends on the condensate density, which is reduced due to the presence of the non-condensate \cite{npp_book}.


\section{\label{sec:stab}3D Stability}

A repercussion of attractive interactions between particles is that, for sufficiently large number of particles and/or interaction strength, the mean-field wave function of a 3D condensate is unstable to collapse. For systems possessing only short-ranged isotropic van der Waals type interactions, the critical point of collapse has been extensively studied in Refs.~\cite{malomed_2007,collapse,parker_2007}, and gives insight into the parameter regimes where one can expect stable soliton dynamics \cite{carr_2002}. The presence of dipole-dipole interactions are expected to modify the collapse point significantly \cite{bohn_2009}, which has recently been explored for lower dimensional systems in Refs.~\cite{eichler_2011,lakomy_2012,baillie_2015} as well as for two-dimensional dipolar bright solitons using 3D simulations \cite{koberle_2012,adhikari_2014}. As well as this, the presence of the dipolar interaction can lead to the spectacular $d$-wave collapse of the condensate in three dimensions \cite{lahaye_2008}. Understanding when the dipolar bright soliton is unstable to collapse in turn allows us to identify regimes of stability applicable to the quasi-one-dimensional dynamics described earlier in Sec.~\ref{sec:coll}.

\subsection{Gaussian Variational Approach}

We employ a variational approach that approximates the wave function of the dipolar soliton as a 3D Gaussian packet with variable width in each dimension \cite{collapse}. This approach has provided important insight into the stability of non-dipolar bright solitons, predicting thresholds for instability which agree closely with experiments \cite{parker_2007}. Under general conditions an appropriate variational ansatz is given by
\begin{equation}\label{eqn:vpsi}
\psi({\bf r}){=}\sqrt{\frac{N}{{\pi}^{\frac{3}{2}}\sigma_x \sigma_y \sigma_z\bar{l}^{3}}}\exp\bigg[{-}\frac{1}{2\bar{l}^{2}}\bigg(\frac{x^2}{\sigma_{x}^{2}}{+}\frac{y^2}{\sigma_{y}^{2}}{+}\frac{z^2}{\sigma_{z}^{2}}\bigg)\bigg]
\end{equation}      
\begin{figure}[t]
\includegraphics[scale=0.8]{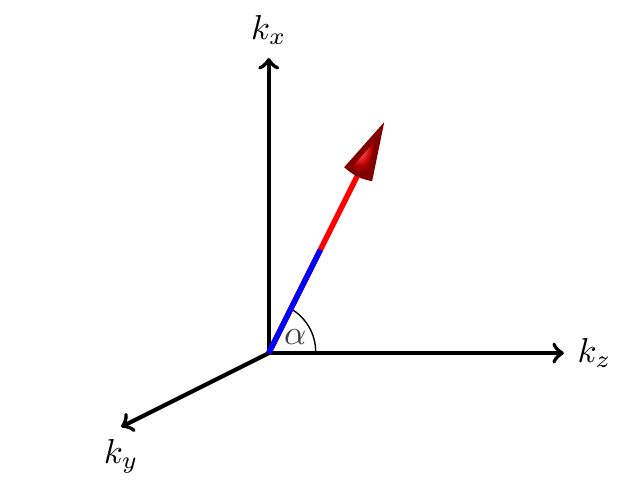}
\caption{\label{fig:dipole_axis}Schematic of the dipoles polarization with respect to the cartesian momentum axis. The polarized dipoles make an angle $\alpha$ with the $k_z$ axis.}
\end{figure}

\noindent where {the lengthscale $\bar{l}=\sqrt{\hbar/m\bar{\omega}}$ is based on the geometric mean of the transverse trapping frequencies $\bar{\omega}=(\omega_x\omega_y)^{1/2}$ and $\sigma_{x,y,z}$ are the dimensionless variational widths of the packet. Equation \eqref{eqn:vpsi} is normalized to the total number of atoms, $\int d{\bf r}|\psi({\bf r})|^2=N$. }

We seek to calculate the total energy of the packet in terms of the above parameters.  We write the total energy as $E=E_{0}+E_{\rm dd}$, where $E_{\rm dd}$ is the dipolar interaction energy, while
\begin{equation}\label{eqn:enz}
E_{0}=\int d{\bf r}\bigg[\frac{\hbar^2}{2m}|\nabla\psi|^2+V(x,y)|\psi|^2+\frac{g}{2}|\psi|^4\bigg]
\end{equation} 
constitutes the remaining energy contributions arising from kinetic energy, potential energy (from the transverse trapping $V(x,y)=\frac{1}{2}m(\omega_{x}^{2}x^2+\omega_{y}^{2}y^2)$) and van der Waals interaction energy.  These contributions to the energy are handled in real space.  Meanwhile, the dipolar contribution $E_{\rm dd}$ is evaluated in momentum space, using the convolution theorem. We consider the case where the atoms forming the condensate are polarized by an external magnetic field such that their individual dipole moments form an angle $\alpha$ with the $z$ axis, as shown schematically in Fig.~\ref{fig:dipole_axis}. This configuration leads to a momentum space pseudo-potential
\begin{equation}\label{eqn:uddk}
\tilde{U}_{\rm dd}({\bf k})=\frac{C_{\rm dd}}{3}\bigg[3\frac{(k_{x}\sin\alpha+k_z\cos\alpha)^2}{k_{x}^{2}+k_{y}^{2}+k_{z}^{2}}-1\bigg],
\end{equation}
where $k_i$ is the component of the momentum in the $i^{\rm th}$ coordinate direction.  The dipolar interaction energy is then \cite{lewenstein_2012}
\begin{equation}\label{eqn:endd}
E_{\rm dd}=\frac{1}{2}\frac{1}{(2\pi)^3}\int {\rm d}^{3}{\bf k}\ {\tilde{U}}_{\rm dd}({\bf k}) {\tilde{n}}({\bf k}) {\tilde{n}}(-{\bf k}).
\end{equation}
We perform the integrations appearing in Eq.~\eqref{eqn:endd} in spherical polar coordinates, and assume the dipoles are polarized parallel to the $z$ axis of the condensate so that $\alpha=0$. Then, using Eqs.~\eqref{eqn:vpsi}-\eqref{eqn:endd}, we can write a general expression for the variational energy of the system as
\begin{figure*}[t]
\includegraphics[scale=0.7]{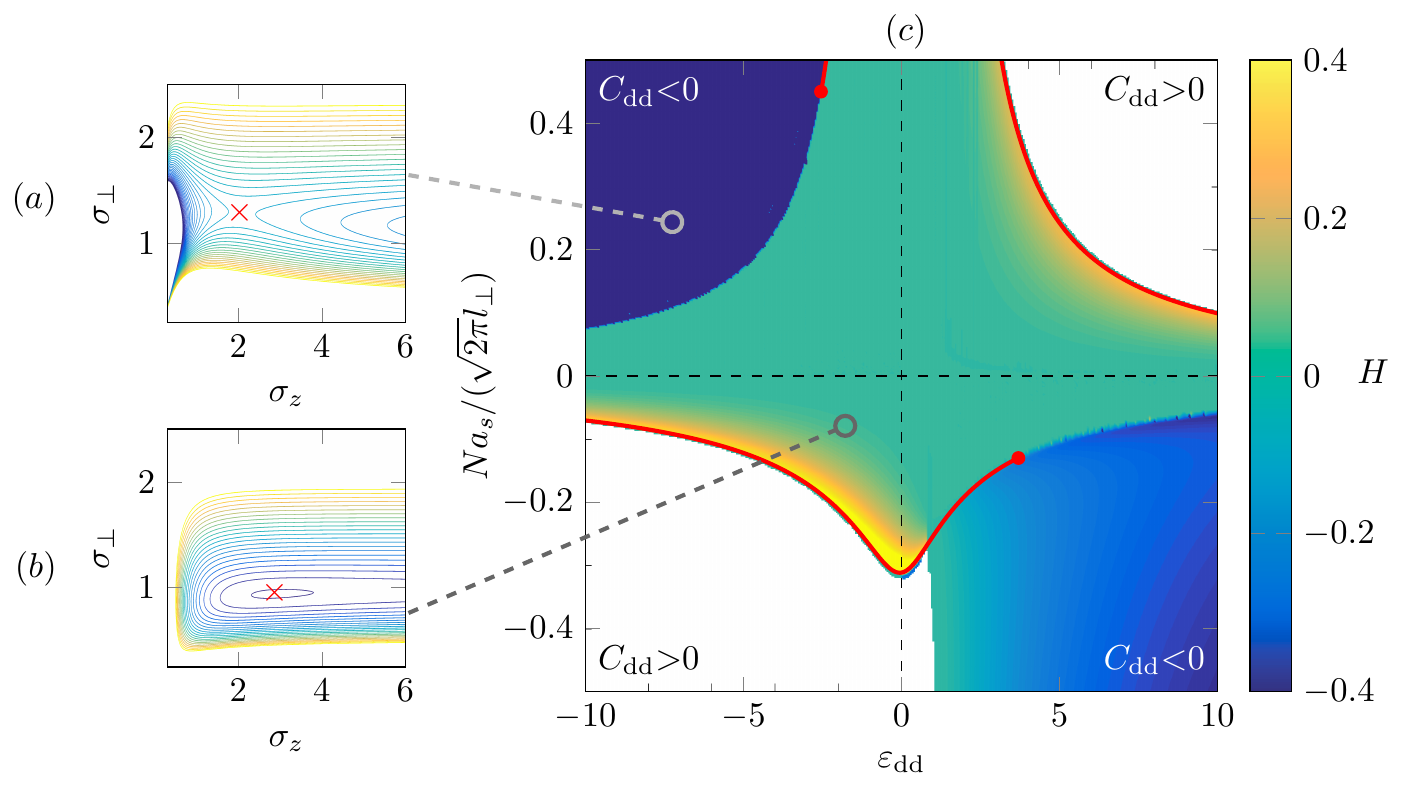}
\includegraphics[scale=0.7]{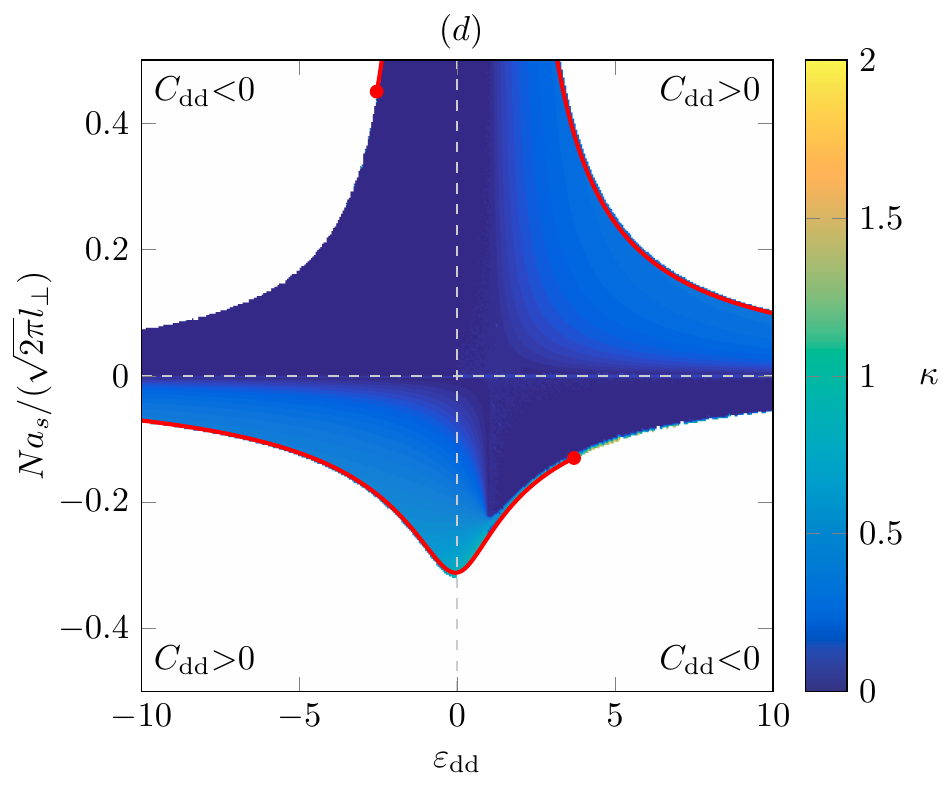}
\caption{\label{fig:stab} Stability analysis of the three-dimensional dipolar soliton. (a, b) Typical plots of the variational energy landscape (corresponding to the parameters annotated in plot (c)). (c) Stability diagram in the ($\varepsilon_{\rm dd}-\beta$) plane. White regions correspond to an absence of a stationary point in the variational energy landscape. Elsewhere, a stationary point exists, and the color denotes the value of the Hessian at that point.  The solid red lines indicate the threshold at which the soliton becomes unstable to collapse. (d) Stability diagram, which in white regions correspond to unstable states and colored regions to stable states; in the latter case the color indicates the aspect ratio $\kappa$ of the variational solution. The dashed lines in (c) and (d) indicate the axes $\varepsilon_{\rm dd}=0$ and $\beta=0$.}
\end{figure*}
\begin{widetext}
\begin{subequations}
\begin{equation}\label{eqn:3dvaren}
\frac{E}{N\hbar\bar{\omega}}=\frac{1}{4}\bigg[\frac{1}{\sigma_{x}^{2}}+\frac{1}{\sigma_{y}^{2}}+\frac{1}{\sigma_{z}^{2}}\bigg]+\frac{1}{4\lambda}\bigg[\sigma_{x}^{2}+\lambda^{2}\sigma_{y}^{2}\bigg]+\frac{\beta}{\sigma_{x}\sigma_{y}\sigma_{z}}\bigg[1-\varepsilon_{\text{dd}}\int\limits^{2\pi}_{0}\frac{{\rm d}\phi}{2\pi}f[\kappa(\phi)]\bigg],
\end{equation}
where $\phi$ denotes the azimuthal angles and $f[\kappa(\phi)]$ is defined as
\begin{equation}\label{eqn:kappafn}
f[\kappa(\phi)]=\frac{1+2\kappa^{2}(\phi)}{1-\kappa^2(\phi)}\frac{\kappa_x\kappa_y}{\kappa^2(\phi)}-3\kappa_x\kappa_y\frac{\text{atanh}\sqrt{1-\kappa^2(\phi)}}{(1-\kappa^2(\phi))^{3/2}}
\end{equation}
\end{subequations}
\end{widetext}
and $\kappa^{2}(\phi)=(\kappa_{x}^{2}-\kappa_{y}^{2})\cos^{2}\phi+\kappa_{y}^{2}$ defines the anisotropic aspect ratio function. Equations \eqref{eqn:3dvaren} and \eqref{eqn:kappafn} also introduce the trap aspect ratio $\lambda=\omega_y/\omega_x$, the dimensionless interaction parameter $\beta=Na_{s}/(l_{\perp}\sqrt{2\pi})$ and the variational aspect ratios $\kappa_{x,y}=\sigma_{x,y}/\sigma_z$. 

\subsection{Stability Analysis}


For a given set of system parameters ($\beta, \varepsilon_{\rm dd}$ and $\lambda$), Eq. (\ref{eqn:3dvaren}) defines an energy landscape as a function of the dimensionless length scales $\sigma_x, \sigma_y$ and $\sigma_z$.  A stable variational solution is an energy minimum (local or global minimum) in this landscape. It is instructive to consider the non-dipolar case as a pedagogical example, which has been studied numerically and using a variational approach \cite{parker_2007,malomed_2007} to determine the parameter regimes where the bright soliton is stable to collapse.
For moderate attractive van der Waals interactions ($N |a_s|/l_\perp \lappeq 0.7$), the energy landscape has a global minimum at the origin ($\sigma_z=\sigma_\perp=0$), representing a collapsed state, and a local minimum at finite $\sigma_z$ and $\sigma_\perp$, representing the 3D bright soliton solution.  This local energy minimum is preserved by a delicate balance between the kinetic and interaction energy.  A saddle point separates the local and global minimum.  For $N |a_s|/l_\perp \approx 0.7$ the saddle point and local minimum merge; this marks the threshold at which the 3D bright solitons become unstable to a runaway collapse.  Meanwhile, for repulsive van der Waals interactions, the energy landscape decreases monotonically towards $\sigma_z \rightarrow \infty$; any wavepacket will disperse axially and no stable solutions exist (unless axial trapping is applied, which we do not considered here). It is the goal of this section to obtain and analyse the nature of the critical points of the full 3D dipolar bright soliton as a function of the interaction parameters $\beta$ and $\varepsilon_{\rm dd}$.


We seek the points in the energy landscape defined by Eq.~\eqref{eqn:3dvaren} at which an instability manifests. This analysis in principle requires us to solve the set of equations defined by $\partial E/\partial\sigma_{x,y,z}=0$ simultaneously with the determinant of the Hessian matrix set equal to zero \cite{carr_2002}, which are in general a set of four coupled algebraic equations in four unknowns. In order to simplify the general Eqs.~\eqref{eqn:3dvaren} and \eqref{eqn:kappafn} but still gain useful insight, we consider the axially-symmetric case for which $\omega_x{=}\omega_y$ and $\sigma_{\perp}{=}\sigma_i$ for $i \in \{x,y\}$. The integral over the azimuthal angle $\phi$ appearing in Eq.~\eqref{eqn:3dvaren} can then be evaluated as the $\phi$ dependency of $\kappa^{2}(\phi)$ is removed in this limit. Although we have assumed the bright soliton is well-described by a Gaussian wave function (Eq.~\eqref{eqn:vpsi}), it is worth contrasting the value at which the instability manifests found from the equivalent analysis for $\varepsilon_{\text{dd}}=0$ assuming either a sech or Gaussian like variational wave function. If $\beta^{*}$ denotes the dimensionless critical collapse parameter, then one finds that ${\beta^{*}_{s}/\beta^{*}_{g}}=\sqrt{{3/\pi}}$, which means that the difference between the two approaches is $\sim$2\%, supporting our choice of a Gaussian variational ansatz. Proceeding, we wish to solve $\partial_{\sigma_{z}}E=0$, $\partial_{\sigma_{\perp}}E=0$ and $H=\text{det}({\bf J}(\nabla E))\equiv 0$, where $\nabla=(\partial_{\sigma_z},\partial_{\sigma_\perp})$, and the matrix elements of the Jacobian ${\bf J}$ are defined as
\begin{equation}\label{eqn:cpeqs1}
J_{ij}=\frac{\partial E}{\partial\sigma_{j}}
\end{equation}
in the space $\{\sigma_{z},\sigma_{\perp}\}$. The value of the Hessian $H$ determines the nature of the extrema. For $H<0$ the extrema is a saddle point; for $H>0$ it is necessary to evaluate either $\partial^{2}_{\sigma_{i}}E$ at the critical point to determine if the extrema is a minimum ($\partial^{2}_{\sigma_{i}}E>0$) or a maximum ($\partial^{2}_{\sigma_{i}}E<0$).

We employ an iterative procedure to obtain numerical solutions for the collapse point in the $(\beta,\varepsilon_{\text{dd}})$ parameter space, using the exact analytical result for $\varepsilon_{\text{dd}}=0$ at which the bright soliton collapses 
as the initial point to start the numerical calculation from. Figure \ref{fig:stab} (c) shows the collapse points (solid red), computed numerically. The coloring of each of the four quadrants is found from the Hessian $H$, evaluated from Eq.~\eqref{eqn:cpeqs1} for a given value of $\beta$ and $\varepsilon_{\rm dd}$, evaluated at the extrema point. The Hessian gives us insight into the nature of the extrema close to and away from the collapse points of the bright soliton. The white areas of this figure are where there are no extrema, and the bright soliton is unstable to collapse.

The collapse dynamics can be categorised based on the sign of the dipole-dipole interaction parameter $C_{\rm dd}$. For $C_{\rm dd}>0$ (bottom left and top right quadrants) the dipoles attract each other in a head-to-tail configuration, and beyond the critical point of collapse no stable dipolar bright soliton can exist (white areas, Fig.~\ref{fig:stab} (c)). On the other hand, when the Hessian $H<0$, there is a saddle point in the energy surface, and no stable bright soliton can form. These regions are found in both the $C_{\rm dd}<0$ parts of the parameter space. Alternatively for $C_{\rm dd}<0$, (bottom right and top left quadrants) the dipoles repel each other in the head-to-tail arrangement, which precludes the bright soliton from collapsing. Instead, one can have a runaway expansion, where the repulsive nature of the dipoles overcomes any attractive forces present. Interestingly, the solutions in these two quadrants do not continue indefinitely, but rather terminate at a point, as indicated by red circles. 

Contour plots of the energy surfaces found from Eqs.~\eqref{eqn:3dvaren}-\eqref{eqn:kappafn} in the $\sigma_z,\sigma_{\perp}$ parameter space accompany Fig.~\ref{fig:stab} (c), Fig.~\ref{fig:stab} (a) and (b). Figure \ref{fig:stab} (a) shows a typical energy contour in a region where the Hessian $H<0$, i.e. a saddle point. Meanwhile, a typical `bowl' configuration that the energy takes in stable regions of the $(\beta,\varepsilon_{\rm dd})$ parameter space is shown in Fig.~\ref{fig:stab} (b). This particular plot shows a stable minima indicative of regions where the bright soliton is stable. Figure \ref{fig:stab} (d) shows a shaded plot of the stable regions where one can expect a bright soliton to form. In comparison with Fig.~\ref{fig:stab} (c), the two regions with $H<0$ ($C_{\rm dd}<0$) have been removed, showing how the stable regions are bound by the collapse or runaway expansion curves (solid red). For $C_{\rm dd}>0$ condensation occurs with $\kappa\lesssim 1$, giving a three-dimensional character to the solitons in these regions of the parameter space. For $C_{\rm dd}<0$ however, one finds instead that $\kappa\ll 1$, indicative of a very elongated (cigar) like cloud. Here the solitons exist closer to the one-dimensional limit. Finally, we note that close to the unstable boundary (red lines) the bright soliton can pass through a region where $\kappa>1$. Such states would be challenging to observe, as the the system would preferentially wish to collapse due to the presence of thermal or quantum fluctuations.

Although our presented results only consider a particular choice of dipole polarization, one can still comment on the stability of the dipolar bright soliton when the dipoles are, say, polarized perpendicular to the axis of the waveguide.  As was noted in Sec.~\ref{sec:dbs}, altering the polarization of the dipoles from $\theta=0$ to $\theta=\pi/2$ has the effect of swapping the regimes of $\varepsilon_{\rm dd}$ where one obtains bright soliton solutions. We can speculate that a similar effect would occur when examining the stability of the dipolar system, except here we would see the regions associated with collapse and runaway expansion switch. However, this case breaks cylindrical symmetry, requiring a fully anisotropic ansatz to capture the stability of the system. This greatly complicates the analysis, as one has three variational width parameters to consider.

\section{Conclusions\label{sec:con}}
We have analysed the solutions, quasi-one-dimensional dynamics and full three-dimensional stability of dipolar bright solitons. The bright soliton solutions obtained from the dipolar Gross-Pitaevskii equation exhibit a number of novel features, including collisions which have regimes of elastic behavior, bound state formation and soliton fusion. These regimes where shown to depend sensitively on the dipolar interactions and the presence of noise, which modify the phase shifts of the solitons. We quantified the collisional behaviour in terms of the coefficient of restitution. Analysis of soliton dynamics in terms of the coefficient of restitution could then provide important insight for systems where the full scattering phase shifts may be difficult to obtain analytically.


The stability of the full three-dimensional dipolar system was explored; in particular it emerged that the dipolar interactions can destabilize the bright soliton in two distinct ways, either through a traditional collapse, or instead through a runaway expansion along the axis. For axially-polarized dipoles the former occurs when the dipole-dipole interaction is positive, while the later is associated with regimes of anti-dipoles for which the dipole-dipole interaction is instead negative.  

Our results provide a benchmark for future experimental studies of nonlocal soliton in dipolar condensates. In turn, this system offers unique opportunities to explore the fundamental properties of nonlocal solitons in general with the immense tunability of atomic physics.

Data supporting this publication is openly available under an Open Data Commons Open Database License \cite{data}.      

\section{Acknowledgements}
M.J.E and N.G.P. acknowledge support from EPSRC (UK) Grant No. EP/M005127/1. T.B. acknowledges support from EPSRC (UK). R. D. thanks Newcastle University for a Vacation Research Scholarship.


\begin{thebibliography}{10}
\bibitem{mollenauer_2006}
L. F. Mollenauer and J. P. Gordon, {\it Solitons in Optical Fibers} (Boston, MA: Acadamic, 2006).
\bibitem{dauxois_2006}
T. Dauxois and M. Peyrard, {\it Physics of Solitons} (Cambridge University Press, Cambridge, 2006).
\bibitem{bloch_2008}
I. Bloch, J. Dalibard, and W. Zweger, {\it Rev. Mod. Phys.} {\bf 88}, 885 (2008).

\bibitem{frantzeskakis_2010} D.~J. Frantzeskakis, {\em J. Phys. A} {\bf 43}, 213001  (2010).
\bibitem{khaykovich_2002}
L. Khaykovich, F. Schreck, G. Ferrari, T. Bourdel, J. Cubizolles, L. D. Carr, Y. Castin, and C. Salomon, {\it Science} {\bf 17}, 1290 (2002).
\bibitem{strecker_2002}
K. E. Strecker, G. B. Partridge, A. G. Truscott, and R. G. Hulet, {\it Nature} {\bf 417}, 150 (2002).
\bibitem{cornish_2006}
S. L. Cornish, S. T. Thompson, and C. E. Wieman, {\it Phys. Rev. Lett.} {\bf 96}, 170401 (2006).
\bibitem{medley_2014}
P. Medley, M. A. Minar, N. C. Cizek, D. Berryrieser, and M. A. Kasevich, {\it Phys. Rev. Lett.} {\bf 112}, 060401 (2014).
\bibitem{mcdonald_2014}
G. D. McDonald, C. C. N. Kuhn, K. S. Hardman, S. Bennetts, P. J. Everitt, P. A. Altin, J. E. Debs, J. D. Close, and N. P. Robins, {\it Phys. Rev. Lett.} {\bf 113}, 013002 (2014).
\bibitem{lepoutre_2016}
S Lepoutre, L Fouch\'e, A Boiss\'e, G Berthet, G Salomon, A Aspect, and T Bourdel, {\it Phys. Rev. A} {\bf 94}, 053626 (2016). 
\bibitem{marchant_2013}
A. L. Marchant, T. P. Billam, T. P. Wiles, M. M. H. Yu, S. A. Gardiner, S. L. Cornish, {\it Nat. Comm.} {\bf 4}, 1865 (2013).
\bibitem{marchant_2016}
A. L. Marchant, T. P. Billam, M. M. H. Yu, A. Rakonjac, J. L. Helm, J. Polo, C. Weiss, S. A. Gardiner, and S. L. Cornish, {\it Phys. Rev. A} {\bf 93}, 021604(R) (2016).
\bibitem{nguyen_2014}
J. H. V. Nguyen, P. Dyke, D. Luo, B. A. Malomed, and R. G. Hulet, {\it Nat. Phys.} {\bf 10}, 918 (2014).
\bibitem{helm_2015}
J. L. Helm, S. L. Cornish, and S. A. Gardiner, {\it Phys. Rev. Lett.} {\bf 114}, {134101} (2015).
\bibitem{gordon_1983}
J. P. Gordon, {\it Opt. Lett.} {\bf 8}, 596 (1983).
\bibitem{scharf_1992}
R. Scharf and A. R. Bishop, {\it Phys. Rev. A} {\bf 46}, R2973(R) (1992).
\bibitem{martin_2007}
A. D. Martin, C. S. Adams, and S. A. Gardiner, {\it Phys. Rev. Lett.} {\bf 98}, 020402 (2007); {\it Phys. Rev. A} {\bf 77}, 013620 (2008).
\bibitem{martin_2016}
A. D. Martin, {\it Phys. Rev. A} {\bf 93}, 023631 (2016).

\bibitem{baizakov_2004}
B. B. Baizakov, B. A. Malomed, and M. Salerno, {\it Phys. Rev. A} {\bf 70}, 053613 (2004).
\bibitem{parker_2008}
N. G. Parker, A. M. Martin, S. L. Cornish, and C. S. Adams, {\it J. Phys. B: At. Mol. Opt. Phys.} {\bf 41}, 045303 (2008).
\bibitem{parker_2009}
N. G. Parker, A. M. Martin, C. S. Adams, S. L. Cornish, {\it Physica D} {\bf 238}, 1456 (2009).
\bibitem{khawaja_2002}
U. Al Khawaja, H. T. C. Stoof, R. G. Hulet, K. E. Strecker, and G. B. Partridge, {\it Phys. Rev. Lett.} {\bf 89}, (2002).
\bibitem{billam_2011}
T. P. Billam, S. L. Cornish, and S. A. Gardiner, {\it Phys. Rev. A} {\bf 83}, 041602(R) (2011).
\bibitem{khawaja_2011}
U. Al Khawaja and H. T. C. Stoof, {\it New. J. Phys.} {\bf 13}, 085003 (2011). 
\bibitem{griesmaier_2005}
A. Griesmaier, J. Werner, S. Hensler, J. Stuhler, and T. Pfau, {\it Phys. Rev. Lett.} {\bf 94}, 160401 (2005).
\bibitem{beaufils_2008}
Q. Beaufils, R. Chicireanu, T. Zanon, B. Laburthe-Tolra, E. Mar\'echal, L. Vernac, J. C. Keller, and O. Gorceix, {\it Phys. Rev. A} {\bf 77}, 061601(R) (2008).
\bibitem{lu_2011}
M. Lu, N. Q. Burdick, S. H. Youn, and B. L. Lev, {\it Phys. Rev. Lett.} {\bf 107}, 190401 (2011).
\bibitem{tang_2015}
Y. Tang, N. Q. Burdick, K. Baumann, and B. L. Lev, {\it New. J. Phys.} {\bf 17}, 045006 (2015).
\bibitem{aikawa_2012}
K. Aikawa, A. Frisch, M. Mark, S. Baier, A. Rietzler, R. Grimm, and F. Ferlaino, {\it Phys. Rev. Lett.} {\bf 108}, 210401 (2012).
\bibitem{lahaye_2009}
T. Lahaye, C. Menotti, L. Santos, M. Lewenstein and T. Pfau, {\it Rep. Prog. Phys.} {\bf 72}, 126401 (2009).
\bibitem{koch_2008}
T. Koch, T. Lahaye, J. Metz, B. Fr\"olich, A. Grismaier, and T. Pfau, {\it Nat. Phys.} {\bf 4}, 218 (2008).
\bibitem{kadau_2016}
H. Kadau, M. Schmitt, M. Wenzel, C. Wink, T. Maier, I. Ferrier-Barbut, and T. Pfau, {\it Nature} {\bf 10}, 1038 (2016).
\bibitem{barbut_2016}
I. F.-Barbut, H. Kadau, M. Schmitt, M. Wenzel, and T. Pfau, {\it Phys. Rev. Lett.} {\bf 116}, 215301 (2016).
\bibitem{schmitt_2016}
M. Schmitt, M. Wenzel, F. B\"ottcher, I. Ferrier-Barbut, and T. Pfau, {\it Nature} {\bf 539}, 259 (2016).
\bibitem{chomaz_2016}
L. Chomaz, S. Baier, D. Petter, M. J. Mark, F. W\"achtler, L. Santos, and F. Ferlaino, {\it Phys. Rev. X} {\bf 6}, 041039 (2016).
\bibitem{baillie_2016}
D. Baillie, R. M. Wilson, R. N. Bisset, and P. B. Blakie, {\it Phys. Rev. A} {\bf 94}, 021602(R) (2016). 
\bibitem{bisset_2016}
R. N. Bisset, R. M. Wilson, D. Baillie, and P. B. Blakie, {\it Phys. Rev. A} {\bf 94}, 033619 (2016).
\bibitem{wachtler_2016}
F. W\"achtler and L. Santos, {\it Phys. Rev. A} {\bf 93}, 061603(R).
\bibitem{wachtler_2016a}
F. W\"achtler and L. Santos, {\it Phys. Rev. A} {\bf 94}, 043618 (2016).
\bibitem{burdick_2016}
N. Q. Burdick, Y. Tang, and B. Lev. {\it Phys. Rev. X} {\bf 6}, 031022 (2016).
\bibitem{rotschild_2006}
C. Rotschild, B. Alfassi, O. Cohen, and M. Segev, {\it Nat. Phys.} {\bf 2}, 769 (2006).
\bibitem{piccardi_2011}
A. Piccardi, A. Alberucci, N. Tabiryan, and G. Assanto, {\it Opt. Lett.} {\bf 36}, 1356 (2011).
\bibitem{jang_2013}
J. K. Jang, M. Erkintalo, S. G. Murdoch and S. Coen, {\it Nat. Photonics} {\bf 7}, 657 (2013).
\bibitem{cuevas_2009}
J. Cuevas, B. Malomed, P. G. Kevrekidis, and D. J. Frantzeskakis, {\it Phys. Rev. A} {\bf 79}, 053608 (2009).
\bibitem{baizakov_2015}
B. B. Baizakov, S. M. Al-Marzoug, and H. Bahlouli, {\it Phys. Rev. A} {\bf 92}, 033605 (2015).
\bibitem{umarov_2016}
B. A. Umarov, N. A. B. Aklan, B. B. Baizakov, and F Kh Abdullaev, {\it J. Phys. B: At. Mol. Opt. Phys.} {\bf 49}, 125307 (2016).
\bibitem{pawlowski_2015}
K. Paw\l{}owski and K. Rz\c{a}\.{z}ewski, {\it New. J. Phys.} {\bf 17}, 105006 (2015).
\bibitem{bland_2015}
T. Bland, M. J. Edmonds, N. P. Proukakis, A. M. Martin, D. H. J. O'Dell, and N. G. Parker, {\it Phys. Rev. A} {\bf 92}, 063601 (2015).
\bibitem{edmonds_2016}
M. J. Edmonds, T. Bland, D. H. J. O'Dell, and N. G. Parker, {\it Phys. Rev. A} {\bf 93}, 063617 (2016).
\bibitem{bland_2016}
T. Bland, K. Paw\l{}owski, M. J. Edmonds, K. Rz\c{a}\.{z}ewski, N. G. Parker, arXiv:1610.02002
\bibitem{pedri_2005}
P. Pedri and L. Santos, {\it Phys. Rev. Lett.} {\bf 95}, 200404 (2005).
\bibitem{tikhonenkov_2008}
I. Tikhonenkov, B. A. Malomed, and A. Vardi, {\it Phys. Rev. Lett.} {\bf 100}, 090406 (2008).
\bibitem{raghunandan_2015}
M. Raghunandan, C. Mishra, K. \L{}akomy, P. Pedri, L. Santos, and R. Nath, {\it Phys. Rev. A} {\bf 92}, 013637 (2015).
\bibitem{chen_2016}
H. Chen, Y. Liu, Q. Zhang, Y. Shi, W. Pang, and Y. Li, {\it Phys. Rev. A} {\bf 93}, 053608 (2016).
\bibitem{giovanazzi_2002}
S. Giovanazzi, A. G\"orlitz, and T. Pfau, {\it Phys. Rev. Lett.} {\bf 89}, 130401 (2002).
\bibitem{santos_2000}
L. Santos, G. V. Shlyapnikov, P. Zoller, and M. Lewenstein, {\it Phys. Rev. Lett.} {\bf 85}, 1791 (2000).
\bibitem{parker_2008a} 
N. G. Parker and D. H. J. O'Dell, {\it Phys. Rev. A} {\bf 78}, 041601(R) (2008).
\bibitem{cai_2010} Y. Cai, M. Rosenkranz, Z. Lei and W. Bao, {\it Phys. Rev. A} 82, 043623 (2010). 

\bibitem{craig_1999}
D. P. Craig and T. Thirunamachandran, {\it Molecular Quantum Electrodynamics} (Dover Publications, London, 1999).
\bibitem{giovanazzi_2004}
S. Giovanazzi and D. H. J. O'Dell, {\it Eur. Phys. J. D} {\bf 31}, 439 (2004).
\bibitem{campbell_1986}
D. K. Campbell, M. Peyrard, and P. Sodano, {\it Physica D} {\bf 19}, 165 (1986).
\bibitem{wuster_2009}
B. J. Dabrowska-W\"uster, S. W\"uster, and M. J. Davis, {\it New. J. Phys.} {\bf 11}, 053017 (2009).
\bibitem{npp_book}
N. P. Proukakis, S. A. Gardiner, M. J. Davis and M. H. Szymanska, {\it Quantum Gases: Finite Temperature and Non-Equilibrium Dynamics} (Imperial College Press, 2013).
\bibitem{collapse} V. M. Perez-Garcia, H. Michinel, J. I. Cirac, M. Lewenstein,
and P. Zoller, {\it Phys. Rev. A} {\bf 56}, 1424 (1997);
V. M. Perez-Garcia, H. Michinel, and H. Herrero, {\it Phys. Rev. A} {\bf 57}, 3837 (1998); A. Gammal, T. Frederico and L. Tomio, {\it Phys. Rev. A} {\bf 64} 055602 (2001); L. Salasnich, {\it Phys. Rev. A} {\bf 70}, 053617 (2004).
\bibitem{malomed_2007}
B. A. Malomed, F. Lederer, D. Mazilu, D. Mihalache, {\it Phys. Lett. A} {\bf 361}, 336 (2007).
\bibitem{parker_2007} N. G. Parker, S. L. Cornish, C. S. Adams and A. M. Martin, {\it J. Phys. B: At. Mol. Opt. Phys.} {\bf 40}, 3127 (2007).

\bibitem{carr_2002}
L. D. Carr and Y. Castin, {\it Phys. Rev. A} {\bf 66}, 063602 (2002).
\bibitem{bohn_2009}
J. L. Bohn, R. M. Wilson, and S. Ronen, {\it Laser Phys.} {\bf 19} 547 (2009).
\bibitem{eichler_2011}
R. Eichler, J\"org Main, and G\"unter Wunner, {\it Phys. Rev. A} {\bf 83}, 053604 (2011).
\bibitem{lakomy_2012}
K. \L{}akomy, Rejish Nath, and Luis Santos, {\it Phys. Rev. A} {\bf 86}, 013610 (2012).
\bibitem{baillie_2015}
D. Baillie and P. B. Blakie, {\it New. J. Phys.} {\bf 17}, 033028 (2015).
\bibitem{koberle_2012}
P. K\"oberle, D. Zajec, G. Wunner, and B. A. Malomed, {\it Phys. Rev. A} {\bf 85}, 023630 (2012).
\bibitem{adhikari_2014}
S. K. Adhikari, {\it Phys. Rev. A} {\bf 90}, 055601 (2014).
\bibitem{lahaye_2008}
T. Lahaye, J. Metz, B. Fr\"ohlich, T. Koch, M. Meister, A. Griesmaier, T. Pfau, H. Saito, Y. Kawaguchi, and M. Ueda, {\it Phys. Rev. Lett.} {\bf 101}, 080401 (2008).
\bibitem{lewenstein_2012}
M. Lewenstein, A. Sanpera, and V. Ahufinger, {\it Ultracold Atoms in Optical Lattices} (Oxford University Press, 2012).
\bibitem{data} Newcastle University Data (doi to be added).

\end{thebibliography}
\end{document}